\documentclass[12pt,a4paper]{article}
\usepackage{psfig,amssymb,latexsym,fancyhdr,graphics}

\title{\bf A New Space-Time Model for Volatility Clustering in the Financial Market}
\author{Boguta, M.\ and J\"arpe, E.}

\newtheorem{lem}{\bf Lemma}
\newtheorem{prop}{\bf Proposition}
\newtheorem{thm}{\bf Theorem}

\newtheorem{ex}{\normalfont\bfseries Example}

\begin{document}
\maketitle
\begin{abstract}
A new space-time model for interacting agents on the financial market is
presented. It is a combination of the Curie-Weiss model and a space-time
model introduced by J\"arpe~\cite{jarpe1}. Properties of the model are
derived with focus on the critical temperature and magnetization. It turns
out that the Hamiltonian is a sufficient for the temperature parameter and
thus statstical inference about this parameter can be performed. Thus e.g.\
statements about how far the current financial situation is from a financial
crisis can be made, and financial trading stability be monitored for
detection of malicious risk indicating signals.
\end{abstract}

\section{Introduction}
\label{sec_intr}
The foundation of the Ising model became an very important event in the modern
physics. It was the basic tool explaining critical temperatures for which phase
transitions occur in physical systems (see Domb et al~\cite{domb_et_al}). It
is one of the most studied models with wide range of applications in different
sciences. Weidlich used Ising model to explain the polarization phenomena in
sociology. It also has been adapted in economics to explain the diffusion
of technical innovations. New technologies were stated as the result of the
interaction with neighboring firms. Into general equilibrium economics Ising
model was introduced by F\"ollmer~\cite{follmer}. There are several sources
of influence on the price of the stock. One important component is real firm
data; another one, correlation between amount of buyers and sellers. The
action of each trader (he is a buyer or seller) was taken as the value of
the trader. Kaizoji~\cite{kaizoji} introduced an Ising-type model of
speculative activity, which explain bubbles and crashes in stock market. He
introduced the market-maker, who adjusts the price on the market in
dependence of correlation of the buyers and sellers.

After Curie had discovered the critical temperature, Weiss developed a theory
of ferromagnetism based on a spin system. It appears by replacement of the
nearest-neighbor pairs interacting of the Ising model by assumption that
each spin variable interacts with each other spin variable at any site of the
lattice with exactly the same strength.

Some space-time models have been suggested. One model based on the Ising model,
was suggested by J\"arpe~\cite{jarpe1}. The partition function of the model
contains two Hamiltonians, one of which describes previous moment of the time.
This model may be used for describing volatility clustering on the market.

The purpose of this paper is to develop a new space-time model, which is also
describes the volatility clustering without assumptions about the structure
of the lattice. Today information about trading on the markets is available,
e.g.\ in Internet, to everyone. Since there is less space restrictions for
this reason it motivates creating a space-time model baced on the Curie-Weiss
model. Such a model is formally defined in this paper, and some results about
critical temperature of the market is derived according to the distribution of
the Hamiltonian in this model.

In Section~\ref{sec_meth} the model and methods which are used are described.
In Section~\ref{sec_res} all main results are presented. The implications are
discussed in Section~\ref{sec_disc}.
%The proofs of the theorems are deferred to the Appendix.

\section{Model and Methods}
\label{sec_meth}
The model which is introduced in this paper
is based on two models. The first is the Curie-Weiss model, which is a simple
modification of the Ising model. It allows all agents in the system to interact
with each other with a constant strength. The second model is the spatio-temporal
model of J\"arpe~\cite{jarpe1} which possesses both spatial and time
dependence. The state of a site in a lattice is depending on the states of
its nearest neighbors and on the global degree of clustering of the previous
pattern.

We took from the Curie-Weiss model the idea of the global interaction and
from the model of J\"arpe the structure of the partition function and obtained
a new space-time model which is appropriate for describing the volatility clustering on
the market.

Let us consider a market which contains $N$ traders symbolically denoted by
$i=1,2,\ldots,N$. In this simple model every trader in a time-period can buy a
fixed amount of stock or sell the same amount. In the first case the "Trader's
decision" is $X_i=1$, in the second case $X_i=-1$. Then $X=(X_1,X_2,\ldots,X_N)$
represents the investment attitude of the market. All traders are neighbors.
That means that each trader knows about the "Trader's decisions" of all others
traders, so his decision is under influence of the others.

A configuration of the model is a specification of "Trader's decisions" of all
traders of the market. With each configuration $x=\{ x_i:i=1,2,\ldots,N\}$
a Hamiltonian or interaction energy,
\[ H(x)\;=\;-\frac 1N\sum_{i=1}^{N-1}\sum_{j=i+1}^N x_ix_j \]
is connected. We will consider this Hamltonian without investment environment.

Let $p(x^k)=P(X=x^k)$ be the probability of observing the state $x^k$ where
$(x^1,x^2,\ldots,x^{2^N})$ is an enumeration of the distinct states of $X$.
Obviously we have that $0\leq p(x^k)\leq 1$ and $\sum_{k=1}^{2^N}p(x^k)=1$.
Further we assume that all states $x$ are possible (i.e $0<p(x^k )<1$).

Now, wanting to minimize the entropy we have an optimization problem of
minimizing
\begin{equation}
\label{opt_problem}
E_P(H(x))\;=\;\sum_{k=1}^{2^N}p(x^k)H(x^k)
\end{equation}
with respect to measure $P$.

Suppose that the energy of each configuration $x_i$ has been determined. The
probability, $P$, that the system has configuration $x$ with energy $H$ if the
configuration at the privious moment of time is given is:
\[ p(x|x')\;=\;P(X_t=x\,|\,X_{t-1}=x')\;=\;Z_{x'}^{-1}\exp\big({\textstyle
  \frac{\beta}{N}}H(x)H(x')\big) \]
where $Z_{x'}=\sum_xexp(\frac\beta N H(x)H(x'))$ is a partition function and
$\beta$ is the market temperature describing the strenght of interaction
between the traders.

\subsection{Sufficient statistic}
\begin{prop}
The statistic $H(X_t)$ is minimal sufficient for inference about the
temperature paramter conditional on the previous state, $H(X_{t-1})$.
\end{prop}

\subsection{The Critical Temperature Of The New Time-Space Model}
The behavior of the system in the Curie-Weiss model is described by the equation
\begin{equation}
m\;=\;\tanh m
\end{equation}
where $m(x)=\frac 1N\sum_i x_i$ represents magnetization of the configuration.
This equation allows us to obtain the property of the temperature of the market
and the critical value of the temperature.

We will use the method of Hartmann and Weigt~\cite{hartmann&weigt} to obtain the
equation of the behavior of new model.

\begin{thm}
The equation of the behavior of the system for the new model is
\begin{equation}
m(t)\;=\;\tanh\left(\frac{m(t)m^2(t-1)\beta}{2}\right)
\end{equation}
\end{thm}

\begin{thm}
The critical temperature of the new model is $\beta=2$. For $\beta<2$ the model
$\{X_t:t\in\mathbb{Z}\}$ is stationary in the space-time sense. For $\beta>2$,
the variable $X_t$ is stationary in the spatial sense conditional on $X_{t-1}=
x_{t-1}$ if $|m(t-1)|\leq\sqrt{\frac 2\beta}$. 
\end{thm}

When $N\rightarrow\infty$ the partition function has the form
\[ Z\;\approx\;\exp(-N f(m_0(t),m(t-1)))\sqrt{\frac{2\pi}{Nf''(m_0(t),m(t-1))}} \]
We know the function $f$ and the form of the $f''$ is deduced from
\[  f(m(t),m(t-1))\;=\;\textstyle{\frac{1+m(t)}{2}\ln(\frac{1+m(t)}{2})+
    \frac{1-m(t)}{2}\ln(\frac{1-m(t)}{2})-\frac\beta 4m^2(t)m^2(t-1)} \]
implying that
\[ f(m(t),m(t-1))\;\big|_{m(t)=0}\;=\;\textstyle{\ln\frac 12} \]
\[ f''(m(t),m(t-1))\;\big|_{m(t)=0}\;=\;\textstyle{1-\frac\beta 2m^2(t-1)} \]
Therefore
\[ Z\;\approx\;2^N\sqrt{\frac{2\pi}{N(1-\frac\beta 2 m^2(t-1))}} \]

Since the Hamiltonian $H$ is sufficient for the temperature parameter, we are
interested in obtaining the distribution of the Hamiltonian. Testing for
dependence will make a null hypothisis assuming independence, and thus we
first consider the distribution of $H$ assuming $\beta=0$. We are interested
in analysis of dependence between $X_i$ and $X_j$ for $i,j=1,\ldots,N$.

\begin{thm}
Let the variables $\{X_i:i=1,\ldots,N\}$ be independent of each other and take
their values in $\{-1, 1\}$ with equal probability $\frac 12$. Then all
non-identical pairwise products are independent, i.e.\ $X_iX_j\bot X_kX_l$
if $i=j,k,l$ or $j=i,k,l$ for any dimension $N$.
\end{thm}

In this paper, we consider a model where $a$ and $b$ are decisions of the
traders. For every trader the probability of their decision to 'buy' or 'sell' is
$\frac 12$ unconditional of the other traders.

Now, let us consider the set of values of the mean field $M=\sum_{i=1}^NX_i$.
For this we have the state space
\[ \left\{{\textstyle -1,-1+\frac 2N,-1+\frac 4N,\ldots,1}\right\} \]
meaning that $M$ can be in $N+1$ states. Consider now values of $M^2$. If $N$
is even, then one value of $M$ is 0 and
\[ M^2\in\left\{{\textstyle 1,(1-\frac 2N)^2,(1-\frac 4N)^2,\ldots,0}\right\} \]
which is to say that $M^2$ could be in $\frac N2+1$ different states. If $N$
is odd, then 0 is not a value of $M$ and
\[ M^2\in\left\{{\textstyle 1,(1-\frac 2N)^2,(1-\frac 4N)^2,\ldots,\frac{1}{N^2}}
   \right\} \]
and in that case $M^2$ could be any of $\frac{N+1}2$ states.

Regarding the Hamiltonian $H(X)=-\frac 1N\sum_{i<j}X_iX_j=-\frac N2M^2+\frac 12$
we consequently the statespace of $H$
\[ \left\{{\textstyle\frac{1-N}2,\frac 12-\frac N2(1-\frac 2N)^2,
   \frac 12-\frac N2(1-\frac 4N)^2,\ldots,\frac 12}\right\} \]
in the case when $N$ is even. If $N$ is odd we get
\[ \left\{{\textstyle\frac{1-N}2,\frac 12-\frac N2(1-\frac 2N)^2,
   \frac 12-\frac N2(1-\frac 4N)^2,\ldots,\frac 12-\frac 2N}\right\} \]

\begin{lem}
If $P(X_i=1)=p$ and $\{X_i:i=1,\ldots,N\}$ are independent, then
$\frac{M(N+1)}{2}\in Bin(N,p)$, where $M=\frac 1N\sum_{i=1}^NX_i$.
\end{lem}

\begin{thm}
The distribution of $\sum_{i<j}X_iX_j$ when $\{X_i\}$ are independent is
\[ P\Bigg(\sum_{i<j}X_iX_j=h\Bigg)\;=\;
   \left\{\begin{array}{ll}
      {N\choose N/2}\frac{1}{2^N} & \mbox{if }h=-N/2\\
      {N\choose(\sqrt{2h+N}+N)/2}\frac{1}{2^{N-1}} & \mbox{if }h\neq -N/2\\
   \end{array}\right. \]
\end{thm}

\subsection{Hamiltonian distribution with dependent traders}
Let us now consider the case when the decisions of the traders are not independent.
\begin{thm}
The distribution of $\sum_{i<j}X_iX_j$ when $\{X_i\}$ are not independent is
\[ P\Bigg(\sum_{i<j}X_iX_j=h\Bigg)\;=\;
   \left\{\begin{array}{ll}
      Z^{-1}e^{-\beta/2}{N\choose N/2}\frac{1}{2^N} & \mbox{if }h=-N/2\\
      Z^{-1}e^{h\beta/N}{N\choose(\sqrt{2h+N}+N)/2}\frac{1}{2^{N-1}} & \mbox{if }
      h\neq -N/2\\
   \end{array}\right. \]
\end{thm}

\subsection{Time dependent process}
From now on we consider the space-time process $X=\{X_t:t\in\mathbb{Z}\}$
where $X_t=\{X_{i,t}:i=1,\ldots,N\}$. Then we have a corresponding sequence
of Mean fields, $\{M_t:t\in\mathbb{Z}\}$ where $M_t=\frac 1N\sum_{i=1}^NX_{i,t}$,
and of Hamiltonians, $\{H_t:t\in\mathbb{Z}\}$ where $H_t=-\frac 1N\sum_{i=1}^{N
-1}\sum_{j=i+1}^NX_{i,t}X_{j,t}$.

\section{Results}
\label{sec_res}
\begin{thm}
The sequence of Mean fields, $\{M_t\}$, is a Markov chain with transition
probabilities
\[ P(M_t=m_t\,|\,M_{t-1}=m_{t-1})\;=\;
   {N\choose N(1+m_t)/2}Z^{-1}_{m_{t-1}}\exp\bigg(\frac{\beta(1-Nm_t^2)(1-Nm_{t-1
   }^2)}{4N}\bigg) \]
\end{thm}

\begin{thm}
The sequence of Hamiltonians, $\{H_t\}$, is a Markov chain with transition
probabilities
\[ P(H_t=h_t\,|\,H_{t-1}=h_{t-1})\;=\;
   \left\{\begin{array}{ll}
   {N\choose N/2}Z^{-1}_{h_{t-1}}\exp(-{\textstyle\frac{\beta}{2}}h_{t-1})
     & \mbox{if }h_t=\frac 12\\[1mm]
   2{N\choose(N+\sqrt{N(1-2h_t)})/2}Z^{-1}_{h_{t-1}}\exp({\textstyle\frac{\beta}{N}}h_t
       h_{t-1}) & \mbox{if }h_t\neq\frac 12
   \end{array}\right. \]
\end{thm}

\begin{thm}
The conditional expectation of the Hamiltonian is
\[ E(H_t\,|\,H_{t-1}=h_{t-1})\;=\;{\textstyle\frac{N}{h_{t-1}}}\cdot\frac{d\ln
   Z_{h_{t-1}}}{d\beta} \]
and the conditional variance is
\[ V(H_t\,|\,H_{t-1}=h_{t-1})\;=\;{\textstyle(\frac{N}{h_{t-1}})^2}\cdot\frac{
   d\ln Z_{h_{t-1}}}{d\beta} \]
\end{thm}

\subsection{Asymptotics}
\begin{thm}
In case with independent trander we have for large $N$
\[ P\Bigg(\sum_{i<j}X_iX_j=h\Bigg)\;\approx\;
   \frac{1}{\sqrt{\pi(2h+N)}}\exp\bigg(-\frac{2h+N}{4}\bigg) \]
\end{thm}

\begin{thm}
For large $N$ the conditional expectation of the Hamiltonian is
\[ E(H_t\,|\,H_{t-1}=h_{t-1})\;\approx\;\frac{N(1-2h_{t-1})}{4h_{t-1}(
   N-\frac\beta 2(1-2h_{t-1}))} \]
and the varinace is
\[ V(H_t\,|\,H_{t-1}=h_{t-1})\;\approx\;\frac{N^2(1-2h_{t-1})^2}{8h_{t
   -1}^2(N-\frac\beta 2(1-2h_{t-1}))^2} \]
\end{thm}

\subsection{Stationarity}
The process $\{X_t\}$ is a time-homogeneous Markov chain because
\[ P(X_{t+1}=x\,|\,X_t=x')\;=\;P(X_t=x\,|\,X_{t-1}=x') \]
for all states $x$ and $x'$ and time-points $t$.

\begin{thm}
The process $\{X_t\}$ is time-reversible and has stationary distribution
\[ \pi(x)\;=\;\frac{Z_x}{\sum_xZ_x} \]
where $Z_x=\sum_y\exp(\frac\beta NH(x)H(y))$.
\end{thm}

\subsection{Exact calculations}
\begin{ex}
We obtained the exact distribution of the statistic $\sum_{i<j}X_{i,t}X_{j,t}$.
Now let us calculate this distribution in the case with 10 traders. We wrote a
program in the R language of programming which calculates the probability of all
possible configurations of the system containing $N$ traders, by determining the
value of the Hamiltonian for each configuration and build the matrix {\tt Vec} of
dimensions
\[ 3\,\times\,(\mbox{"the largest possible value of $H$"}-\mbox{"the smallest
   possible value of $H$"}) \]
where the first line contains all possible values of $H$, the second line: the
number of the configurations that leads to the corresponding value in the first
line, the third: probability of corresponding value from the first line. In
Figure~\ref{fig_exact_distr} we can see the distribution of $H$ for $N=10$.
\begin{figure}[htbp]
  \begin{center}
    \leavevmode
    \psfig{file=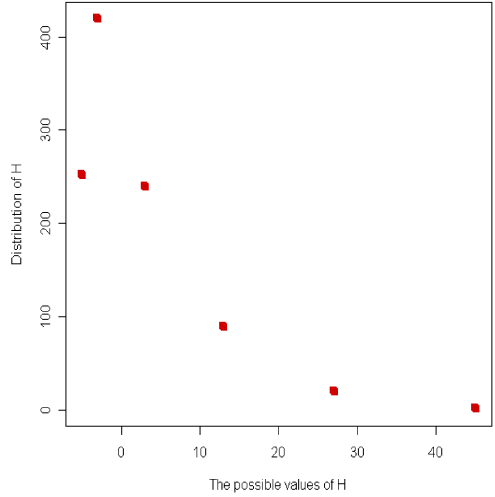,width=65mm}
    \caption{\em Exact distitribution of the energy function when there are 10 traders.}
    \label{fig_exact_distr}
  \end{center}
\end{figure}
In the Table~\ref{table_exact_distr} are the exact numerical values of the
energy distribution.
\begin{table}[htbp]
  \begin{center}
    \begin{tabular}{|l|c|c|c|c|c|c|}\hline
      $h$ & $-5$ & $-3$ & 3 & 13 & 27 & 45\\ \hline
      Number of occurrances & 252 & 420 & 240 & 90 & 20 & 2\\ \hline
      Probability & 0.246 & 0.41 & 0.234 & 0.088 & 0.0195 & 0.00195\\ \hline
    \end{tabular}
    \caption{\em The exact distribution values of the energy function when there are
      10 traders.}
    \label{table_exact_distr}
  \end{center}
\end{table}
\end{ex}

\subsection{Hypothesis test of independence}
If there is only weak interaction between the traders, then the Hamiltonian
is more likely to attain smaller values. If the interactions are stronger,
then larger values of the Hamiltonian is more likely and one talks about
magnetization, which could be an indicator of, or even a cause of, bubbles
and crashes on the market. Therefore methods to state wheter the values of
the observed Hamiltonian is in some dangerous region is of vital importance
to decision makers and inderictly to the whole society.

In order to see if the value of the Hamiltonian deviates from zero to such
an extent that dangerous development is indicated, the correct distribution
is needed. If the Hamiltonian approaches the critical value, the system may
be in a dangerously instable state and a bubble or a crash on the market can
appear. 

Assume that we make $n$ observations $x_1,x_2,\ldots,x_n$ of the system.
Then the corresponding Hamiltonian values $h_1,h_2,\ldots,h_n$ are calculated
and categorized into $K$ classes. If the number of traders are 10 (as in the
previous example), then the we may have 6 classes corresponding to the 6
possible states of the Hamiltonian. But of course we may define these classes
in any way we choose.
% Then we count the number of observations in each class,
% $O_k$ and calculate the expected 

Strong interactions as opposed to near independence is reflected by the
hypotheses
\[ \left\{\begin{array}{l}H_0:\beta=0\\ H_1:\beta>0\end{array}\right. \]
To have argument for dependence, i.e.\ to prove $H_1$, the statistic
\[ S\;=\;\sum_{k=1}^K\frac{(O_k-E_k)^2}{E_k} \]
may be used, where $O_k$ is the number of observations in klass $k$, and
$E_k$ is the expected number of observations in class $k$ according to the
distribution of $S$ under $H_0$ which is $\chi^2_{k-1}$. Thus the null
hypothesis is rejected at level $\alpha$ of significance for values of
$S$ greater than $C$ which is $1-\alpha$ percentile of the $\chi^2$
distribution with $k-1$ degrees of freedom.

\begin{ex}
Let us consider the data from the Swedish steel market for the October 22,
2008. We have the traders and buyers with certain moments of trading. To
analyze these data we will collect information about trading by dividing the
time on parts of each about 10 minutes. We than have the ten most active
traders (AVA, CSB, DBL, ENS, EVL, MSI, NDS, NON, SHB, SWB) and twenty intervals
of activity. Let us consider the Hamiltonian for these traders. The result is
in Table~\ref{tab_indep_test}.
\begin{table}[htbp]
  \begin{center}
  \begin{tabular}{|l|c|c|c|c|}\hline
    State & $-5$ & $-3$ & $3$ & $[13,45]$\\ \hline
    Number of occurances & 2  & 13 & 4 & 1\\ \hline
    Expected number of occurance & 4.92 & 8.2 & 4.688 & 2.19\\ \hline
  \end{tabular}
  \caption{\em The frequencies of observations divided into four classes.}
  \label{tab_indep_test}
  \end{center}
\end{table}
Thus $N=10$ and $K=4$. The value of the test statistic in this case is
\[ {\textstyle\frac{(2-4.92)^2}{4.92}+\frac{(13-8.2)^2}{8,2}+\frac{(4-4.688)^2}{4.688}+
   \frac{(1-2.19)^2}{2.19}}\;=\;5.29\;<\;7,8147\;=\;\chi^2_3 (0.05) \]
% where NH is the number of occurance in the test,
% NE is expected number of occurance.
This means that we can not reject the hypothesis of independence on level 5\%
of significance (or any lower level). As a matter of fact the $p$-value of
$S$ in this case is $0.15175$ so dependence can not be proved on any
reasonable level of significance.
\end{ex}

\section{Discussion}
\label{sec_disc}
In this thesis we investigated a new space-time model for interacting agents
in the financial market. First we reviewed the history of the Ising model
and some other Ising-type models. Then the Ising model, Curie-Weiss model
and some modifications of these models were formally presented. Also we
considered one way of finding the critical temperature of the market. A
new space-time model was developed and necessary and sufficient conditions
for its stationarity were found. The non-linear sensitivity of market global
properties in terms of temperature parameter changes was investigated. The
critical temperature for this model was analytically derived.

The distribution of the Hamiltonian was analyzed using its dependence with
the magnetization of the market and the exact distribution was calculated.
The conditional expectation and variance of the Hamiltonian were found and
the stationary distribution was obtained. Then the exact distribution of
the Hamiltonian for 10 traders was calculated, and the expected distribution
was confirmed. Hypothesis test for independence between agents was considered
for the Swedish steel market and it showed that there is no evident critical
situation on the market at the time of this dataset.

The parameter  reflects how strongly traders are influenced by each other
in the market. It can signalize risk for a crash or a bubble in the market.
Therefore it is very important that its analysis is accurate in such
situations. What remains for future work? We could try a lot of real data
and compare inferential results relying on this model to observable quantities
generally accepted as a measure of health of the situation on the market.
Also we can consider a bigger ammount of traders to have a more exact
$p$-value in the hypothesis testing. Then we can estimate the amount of
interaction for our model using e.g. maximum likelihood estimator. Here
exists also the possibility to develop hypothesis testing based on a time
dependent model. Also interesting to find out how good this model is to
explain volatility clustering.

% \section*{Appendix}
%\setcounter{thm}{0}
%\setcounter{prop}{0}
%\setcounter{lem}{0}
%\setcounter{corr}{0}
%\setcounter{obs}{0}
%\begin{prop}
%The statistic $H(X_t)$ is minimal sufficient for inference about the
%temperature paramter conditional on the previous state, $H(X_{t-1})$.
%\end{prop}

%\pf This is clear from the factorization theorem because the transition
%probability px (x|x ) is a member of the exponential family.
%\epf

%\begin{thm}
%The equation of the behavior of the system for the new model is
%\begin{equation}
%m(t)\;=\;\tanh\left(\frac{m(t)m^2(t-1)\beta}{2}\right)
%\end{equation}
%\end{thm}

%\begin{thm}
%The critical temperature of the new model is $\beta=2$. For $\beta<2$ the model
%$\{X_t:t\in\mathbb{Z}\}$ is stationary in the space-time sense. For $\beta>2$,
%the variable $X_t$ is stationary in the spatial sense conditional on $X_{t-1}=
%x_{t-1}$ if $|m(t-1)|\leq\sqrt{\frac 2\beta}$. 
%\end{thm}

\end{document}